\documentclass[%
 reprint,
nofootinbib,
 amsmath,amssymb,
 aps
]{revtex4-1}
\pdfoutput=1
\usepackage{graphicx}% Include figure files
\usepackage[utf8]{inputenc}
\usepackage{dcolumn}% Align table columns on decimal point
\usepackage{bm}% bold math
\usepackage{verbatim}
\usepackage{color,ulem}

\usepackage[colorlinks = true,
            linkcolor = blue,
            urlcolor  = blue,
            citecolor =green,
            anchorcolor = blue]{hyperref}
\usepackage[utf8]{inputenc}

\usepackage[usenames,dvipsnames]{xcolor}

  \usepackage{titlesec}
\usepackage{chngcntr}
\newcounter{mysection}
\titleclass{\mysection}{straight}[\part]
\titleformat{\mysection}[hang]
  {\normalfont\LARGE\bfseries}{\centering \themysection}{1em}{}
\titlespacing*{\mysection}{0pt}{3.5ex plus 1ex minus .2ex}{2.3 ex plus .2ex}
\renewcommand{\themysection}{\centering \arabic{mysection}}

\begin{document}

\preprint{}[CCTP-2018-3, ITCP-IPP 2017/22]

\title{\huge 
Elasticity bounds from Effective Field Theory
}% Force line breaks with \\

\author{Lasma Alberte}%
 \email{lalberte@ictp.it}
\affiliation{Abdus Salam International Centre for Theoretical Physics (ICTP), Strada Costiera 11, 34151, Trieste, Italy.}%

\author{Matteo Baggioli}%
 \email{mbaggioli@physics.uoc.gr}
 \thanks{\url{www.thegrumpyscientist.com} }
\affiliation{Crete Center for Theoretical Physics, Institute for Theoretical and Computational Physics\\ Department of Physics, University of Crete, 71003
Heraklion, Greece.
}%

\author{Víctor Cáncer Castillo}%
 \email{vcancer@ifae.es} \author{Oriol Pujol{\`a}s}%
 \email{pujolas@ifae.es}
\affiliation{Institut de F\'isica d'Altes Energies (IFAE), The Barcelona Institute of Science and Technology (BIST)\\
Campus UAB, 08193 Bellaterra, Barcelona.
}%

\begin{abstract}

Phonons in solid materials can be understood as the Goldstone bosons of the spontaneously broken spacetime symmetries. 
As such their low energy dynamics are greatly constrained and can be captured by standard effective field theory (EFT) methods. 
In particular, knowledge of the nonlinear stress-strain curves completely fixes the full effective Lagrangian 
at leading order in derivatives. 
We attempt to illustrate the potential of effective methods focusing on the so-called hyperelastic materials, which allow large elastic deformations.
We find that the self-consistency of the EFT imposes a number of bounds on physical quantities, mainly on the maximum strain and maximum stress that can be supported by the medium. 
In particular, for stress-strain relations that at large deformations are characterized by a power-law behaviour $\sigma(\varepsilon)\sim \varepsilon^\nu$, the maximum strain exhibits a sharp correlation with the exponent $\nu$.

\end{abstract}

\pacs{Valid PACS appear here}
\maketitle

\section{Introduction}

A prominent and early example of an Effective Field Theory (EFT) is the theory of elasticity: the continuum-limit description of a solid's mechanical response, including its sound wave excitations -- the phonons \cite{Lubensky,landau7}. 

As in hydrodynamics, elasticity theory can be phrased as a derivative expansion for an effective degree of freedom -- the displacement vector of the solid elements with respect to equilibrium. 
Importantly, the classic elasticity theory can be promoted to the nonlinear regime, addressing  the response to finite deformations \cite{enlighten70333,ZAMM:ZAMM19850650903,Ogden2004}. Operationally, this is done by finding the stress-strain relations for both finite shear or bulk strain applied to the material. These diagrams encode several 
response parameters (such as the {\it proportional limit} or the {\it failure point}, see \cite{Ogden2004} for definitions), which are well defined material properties that go deep into the nonlinear response regime. Typically, these parameters are difficult to compute from the microscopic constituents, so there is a chance that EFT methods may help in understanding some nonlinear elasticity phenomena.

From the viewpoint of quantum field theory (QFT), it is clear that elasticity theory can be treated as a non-trivial ({\it i.e.}, interacting) EFT. 
The way how this theory works as an EFT, however, is quite different from other well known examples, mostly because the underlying symmetry breaking pattern involves spacetime symmetries. 

The purpose of this work is to revisit finite elasticity theory from the viewpoint of QFT. We aim at clarifying
how the EFT methodology works for broken spacetime symmetries and find novel relations between (and bounds on) various nonlinear elasticity parameters.

\section{From Goldstones to stress-strain curves}

%Let us start the discussion 
We start by stating the precise QFT sense in which elasticity theory can be treated as an EFT. The first requirement is that the material must have a separation of scales: we shall consider only low frequency (acoustic) phonons; any other mode is considered as much heavier and integrated-out. (Materials displaying scale invariance violate this assumption and deserve a separate treatment.)
Under this  condition we can exploit the fact that the phonons can be viewed as the Goldstone bosons of translational symmetry breaking~\cite{Leutwyler:1996er,Dubovsky:2005xd,Nicolis:2015sra}. As such we obtain their fully nonlinear effective action by the means of the standard coset construction~\cite{Nicolis:2013lma}.

For simplicity, we shall work in $2+1$ spacetime dimensions, where the 
dynamical degrees of freedom are contained in two scalar fields $\phi^I(x)$.

The internal symmetry group is assumed to be the two-dimensional Euclidean group, $ISO(2)$, acting like translations and rotations in the scalar fields space. The theory then must be shift invariant in the $\phi^I$'s implying that any field configuration  that is linear in the spacetime coordinates will satisfy the equations of motion. 
The equilibrium configuration of an isotropic material is given by:

\begin{equation}\label{equil}
{\phi}^I_{\text{eq}}  =  \delta^I_J \, x^J\,.
\end{equation}
This vacuum expectation value spontaneously breaks the symmetry group $ISO(2)\times ISO(2,1)$ down to the diagonal subgroup. 

Following the coset construction method, one concludes that the effective action at lowest order in  derivatives takes the form
\begin{equation}\label{action}
S= -\int d^3x\,\sqrt{-g}\, V(X,Z)\,,
\end{equation}
with $X$ and $Z$ defined in terms of the scalar fields matrix\footnote{We retain the curved spacetime metric $g_{\mu\nu}$ only to make it clear how the energy-momentum tensor  arises from this action. In practice we shall always work on the Minkowski background $\eta_{\mu\nu} = \textrm{diag}\,(-1,+1,+1)$.} 
$\mathcal I^{IJ}= g^{\mu\nu}\partial_\mu\phi^I\partial_\nu\phi^J$ as 
$
X =\mathrm{tr }\, \big(\mathcal I^{IJ} \big) \,,
\, Z = \det \big(\mathcal I^{IJ} \big)$.
The function $V(X,Z)$ is `free' and its form depends on the solid.
In this language, the phonons $\pi^I$ are identified as the small excitations around the equilibrium configuration defined through $\phi^I= {\phi}^I_{\text{eq}}  +\pi^I$. Plugging this decomposition into \eqref{action} one can find the phonon kinetic terms and their self-interactions  $(\partial\pi)^n$.
The leading phonon effective operators are determined by a few {\it Wilson coefficients} that are related to 
the lowest derivatives of $V$ evaluated on the equilibrium configuration, see \cite{Endlich:2010hf} for details. (Analogous results can be found in \cite{Greiter:1989qb} for superconductors.) The effective action \eqref{action} also encodes the response to finite (large) deformations, and for that  the global form of $V(X,Z)$ is needed.

By symmetry considerations one cannot restrict the action ~\eqref{action} any further. 

To identify what is the function $V(X,Z)$ for a given material one needs more information, some kind of constitutive relation. 
According to the finite elasticity literature (see \textit{e.g.}, \cite{Ogden2004}) the function $V(X,Z)$ is naturally identified with the so-called {\it strain-energy function}. 

This is a function of the  principal invariants characterizing the materials state of deformation. It encodes the full nonlinear response for the so-called Cauchy hyperelastic solids, for which  plastic and  dissipative effects can be ignored \cite{ZAMM:ZAMM19850650903}. 

The form of $V$ can then be found from the stress-strain relations measured in both the shear and the bulk channels of real solids (see, \textit{e.g.} \cite{ZAMM:ZAMM19850650903,enlighten70333,international2002atlas}). More specifically, from the response of the material to constant and homogeneous deformations. 
These can be reduced to configurations of the form
\begin{equation}\label{finitedef}
\phi_{\text{str}}^I=
O^I_Jx^J\;,\;\;
O^I_J\,=\,\alpha\begin{pmatrix} 
\sqrt{1+\varepsilon^2/4} & \varepsilon/2 \\
\varepsilon/2 & \sqrt{1+\varepsilon^2/4} 
\end{pmatrix},
\end{equation}

where $\varepsilon$ and $\alpha-1$ are the shear and the bulk strains respectively, and they induce constant but non-trivial values of $X|_{\text{str}}=\alpha^2(2+\varepsilon^2)$ and $Z|_{\text{str}}=\alpha^4$.
%in our parameterization. 
The amount of stress in the material generated by (or needed to support) such a configuration depends only on the strains $\varepsilon$, $\alpha$ and on the shape of $V(X,Z)$, see {\it e.g.} Eq.~\eqref{sigma}. The upshot is that it is possible to reconstruct the full form of the effective Lagrangian (up to an irrelevant overall constant) by just measuring the stress-strain relations, that is, from the response to time-independent and homogeneous deformations. %
This already illustrates how the solid EFTs retain predictive power.

The next apparent challenge from the QFT viewpoint is that the real world stress-strain curves typically exhibit a dramatic feature:  they terminate at some point, corresponding to the breaking (or elastic failure) of the material. It is then natural to ask how exactly is the  breaking seen in the EFT. Must the function $V(X,Z)$ be singular? Or does the breaking  correspond to a dynamical process ({\it e.g.}, an instability) that can be captured within the EFT %even 
with a regular $V(X,Z)$? We argue below that the latter possibility can certainly arise allowing to extract relations between the parameters that control the large deformations.

\begin{figure}[t]
    \centering
    \includegraphics[width=0.45\textwidth]{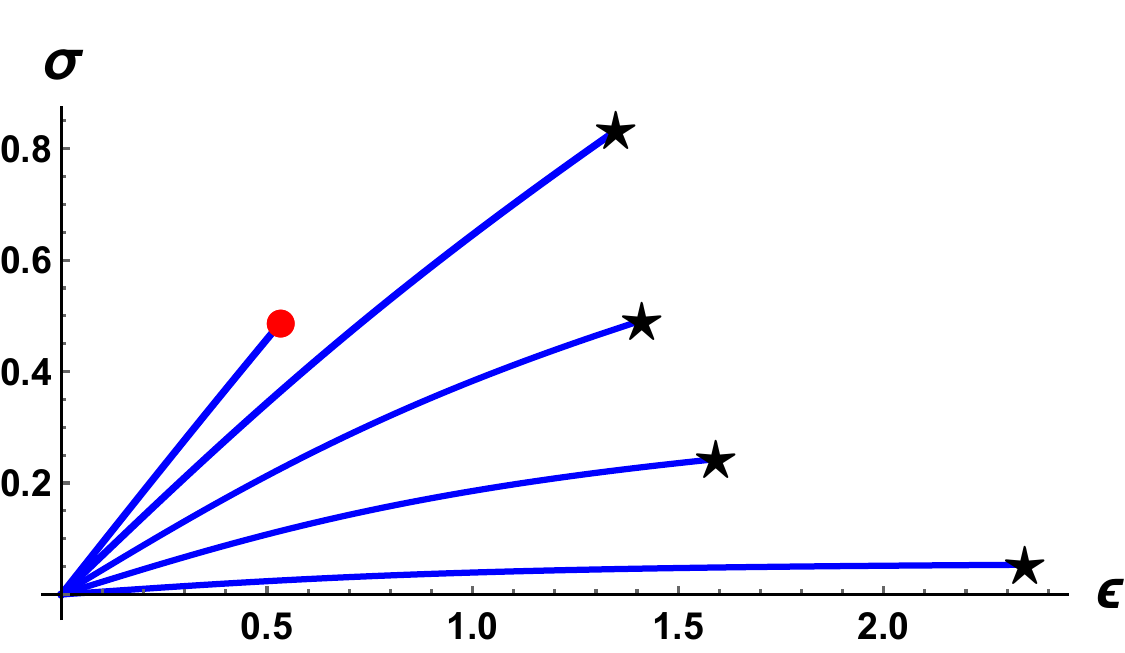}
    \caption{The nonlinear shear stress-strain curve $\sigma(\varepsilon)$ for the benchmark model \eqref{bench} for $B=1.6$ and $A=0.05,0.2,0.35,0.5,0.61$ (from bottom to top). The black stars represent the `breaking' points of the material arising due to the onset of gradient instability; the red dot indicates the onset of superluminality.}
    \label{fig5}
\end{figure}

The main task then is to analyze the stability properties of the strained configuration~\eqref{finitedef}. 
This can be done by setting 
$\phi^I= {\phi}^I_{\text{str}}  +\pi^I$ in \eqref{action} and expanding for `small' $\pi^I$.
In doing so one easily finds that the phonon sound speeds depend on the applied strain $O^I_J$. 
This is a long known phenomenon, the \textit{acoustoelastic effect}, see \textit{e.g.} \cite{Tang1967,PhysRev.92.1145,doi:10.1121/1.426255,doi:10.1121/1.1908623,ABIZA2012364,baba}. 
Still, we argue here that this can have a great impact on the stress-strain relations, eventually  limiting the maximal stress that a material can withstand. 
The reason is that %very 
generically, increasing the strain results into increasing/decreasing the various sound speeds -- typically in an unbounded fashion.
In particular, in most cases past some large enough strain value, $\varepsilon_{\text{max}}$, %one is typically left with one of the following options: 
one of the sound speeds becomes either $i)$  imaginary or $ii)$  superluminal. 
Case $i)$ implies that the material develops an instability and it must evolve to a different ground state. 
Case $ii)$ prevents the existence of a Lorentz invariant ultraviolate completion. Therefore the effective low energy description~\eqref{action} must be physically invalid at least for such a large deformation. 
%\textcolor{red}{cannot be directly linked to the stability of the configuration/material. Instead, it} indicates that, at best, the particular function $V(X,Z)$ for which this happens
%\eqref{bench} 
%misses important ingredients for describing the actual physical system, such as additional degrees of freedom. 
In any  case, one can translate the constraints $i)$~and $ii)$ as upper bounds on the maximum allowed strain that is compatible with the given choice of $V(Z,X)$. 
We remark that these bounds arise even for smooth choices of the effective Lagrangian $V(Z,X)$,
and yet they naturally lead to  stress-strain curves that terminate at some point
$\varepsilon=\varepsilon_{\text{max}}$, see Fig.~\ref{fig5} for some illustrative examples. 

%In any of the two cases, the stress-strain diagram valid for a given $V(X,Z)$ certainly terminates at that maximum strain $\varepsilon_{\text{max}}$. 
%
%(Another possibility is that at large enough strain one of the speeds becomes superluminal. \LA{I don’t really like the following statement. Is it really what we want to say? Oriol? I color it also in a later footnote. Since this would conflict the possibility to embed the material into a Lorentz invariant UV completion, this also leads to physical bounds.})
%
%\OP{This shows how by studying the EFT in the large strain regime can establish relations between the parameters that are involved in the elastic failure for real-world material. }
%
%

Additionally, demanding that none of these pathologies occur for materials that we know admit large deformations (elastomers) 
significantly constrains the stress-strain curves and therefore the possible nonlinear response of materials on quite general grounds. 
We illustrate the point by focusing on materials/EFTs which allow for large deformations {\it and} which realize stress-strain curves with a power-law scaling 
\begin{equation}\label{power}
\sigma\sim\varepsilon^\nu 
\qquad \text{for} \qquad \varepsilon\gg1~.
\end{equation}
Henceforth we shall refer to  $\nu$ as the {\it strain exponent}.
As we show below, both the maximum strain and the exponent $\nu$ are bounded from above, and there is a general relation between the two. It is unclear to us to what extent these results were already known before. Nonetheless, our main goal is to show how the EFT perspective presented here brings some additional layer of understanding to these phenomena.

%Let us summarize here a few important results that immediately follow from the effective action \eqref{action}, still for a generic function $V$.

Firstly, let us obtain the corresponding stress-energy tensor by varying the action with respect to the curved spacetime metric $g_{\mu\nu}$ and evaluating it on the Minkowski background, $g_{\mu\nu}=\eta_{\mu\nu}$:
\begin{align}
T_{\mu\nu}\,=&\,-\,\frac{2}{\sqrt{-g}}\,\frac{\delta S}{\delta g^{\mu\nu}}\,\Big|_{g=\eta}=\,-\,\eta_{\mu\nu}\,V\,+\,2\,\partial_\mu \phi^I \partial_\nu \phi_I\,V_X\,\nonumber \\ &+\,2\,\left(\partial_\mu \phi^I \partial_\nu \phi_I\,X\,-\,\partial_\mu \phi^I \partial_\nu \phi^J \,\mathcal{I}_{IJ}\right)\,V_Z~.
\end{align}
For any time independent scalar field configurations, the stress-energy tensor components are
\begin{align}\label{rho}
&T^{tt}\,\equiv\,\rho\,=\,V\,,\\\label{pressure}
& T^x_x\,\equiv\,\,p\,=\,-\,V\,+\,X\,V_X\,+\,2\,Z\,V_Z\,,\\\label{txy}
&T^x_{y}\,=\,2\,\partial_x \phi^I \partial_y \phi^I\,V_X\,,
\end{align}
where $V_X\equiv\partial V /\partial X,$ etc. Henceforth we shall work with the deformed field configuration \eqref{finitedef} which,  introduces both shear and bulk deformation. In particular, when setting $\alpha=1$, it describes a \textit{pure shear} strain (\textit{i.e.} volume-preserving) in the $(x,y)$ directions induced by $\varepsilon\neq 0$. %On the other hand, 
For $\varepsilon=0$ and $\alpha\neq 1$, the same setup encodes a \textit{pure bulk} strain. In the considered scalar field background configuration $X$ and $Z$ take the values: $ X|_{\text{str}} = \alpha^2(2+\varepsilon^2)\,,\,\,  Z|_{\text{str}} = \alpha^4 $. 

In particular, the full nonlinear stress-strain curve for pure shear deformations as a function of $\varepsilon$ reads:
\begin{equation}\label{sigma}
\sigma(\varepsilon)\equiv T_{xy}=2\varepsilon\sqrt{1+\frac{\varepsilon^2}{4}} \;
V_X\left(2+\varepsilon^2,1\right)\,.
\end{equation}

 The analogous stress-strain curve for pure bulk deformations can also be found  by expressing $\Delta T^x_x= T^x_x-\big.T^x_x\big |_{\text{eq}}$ as a function of the bulk strain, $\alpha-1$. It is thus clear that  from the knowledge (measurement) of both shear and bulk diagrams one can extract the shape of $V(X,Z)$ -- the full effective Lagrangian. For instance, under the assumption that the $Z-$dependence is negligible, then from a given $\sigma(\varepsilon)$ shear stress-strain curve one can extract 
$$
V(X)\simeq \int_2^X dx \;\frac{\sigma(\sqrt{x-2})} {\sqrt{x^2/4\,-\,1}}~.
$$

To make the connection to the linear elasticity theory  explicit one considers small shear and bulk deformations {\it i.e.} small values of $\varepsilon$ and $\alpha-1$. Then, as usual, elastic deformations at linear level are described in terms of the displacement tensor%\footnote{For linear rseponse the distinction between   capital and lowercase indices $I,J,i,j,\dots$ can be dropped}
\begin{equation}\label{eij}
\varepsilon_{ij}\,=\,\frac{1}{2}\left(\partial_i \delta\phi_j\,+\,\partial_j\,\delta\phi_i\right)\,,
\end{equation}
where $\delta\phi^I\equiv\phi^I-\phi^I_{\text{eq}}$ is the displacement away from the equilibrium state, $\phi^I_{\text{eq}} = x^I$.
A deformation of the body that changes its volume is given by the compression or bulk strain as $\varepsilon_{ii}=\,\partial_i\delta\phi^i$. 
In turn, a deformation that only affects its shape -- pure shear -- is given by  $\varepsilon_{ik}-\frac{1}{2}\delta_{ik}\varepsilon_{jj}$.

Expanding both the stress-energy tensor components \eqref{pressure}, \eqref{txy} and the displacement tensor \eqref{eij} up to linear order in $\varepsilon$ and $\alpha-1$ one recovers the usual expression in 2+1 dimensions: ~~~ 
\begin{equation}
T^{\text{lin}}_{ij}\,=\,\left(p+{K}\,\varepsilon_{kk}\right)\delta_{ij}+2\,{G}\left(\varepsilon_{ij}-\frac{1}{2}\delta_{ij}\varepsilon_{kk}\right)\,,\label{str}
\end{equation}
where $p$ is the equilibrium pressure and ${G},{K}$ are the shear and bulk elastic moduli. In the case of a pure shear deformation this gives
$T_{xy}=2G\varepsilon_{xy}+\dots$ and we can read off the shear modulus $G$ as
\begin{equation}\label{Glin}
G= 2\,V_X(2,1)\,.
\end{equation}

Similarly for the case of pure bulk strain ($\varepsilon=0$) we first note that the equation \eqref{pressure} holds at nonlinear level, \textit{i.e.} for arbitrarily large values of $\alpha$. 
In order to find the linear bulk modulus we expand both the bulk strain and the bulk stress $\Delta T^x_x$ around the equilibrium value $\alpha=1$. For the stress this gives $\Delta T_{ii}=2K\varepsilon_{ii}+\dots$ with the equilibrium pressure given in \eqref{pressure} %and evaluated at $\alpha=1$ 
and $\varepsilon_{ii}= 2(\alpha-1)$. The bulk modulus is then: 
\begin{align}\label{Klin}
&{K}\,=\,2ZV_Z+4Z^2 V_{ZZ}+4XZ V_{XZ}+X^2 V_{XX}\,,
\end{align}
where all the quantities are evaluated at $X=2\,,Z=1$. 

All the details concerning the consistency and stability of perturbations around the strained background configuration are given in Appendix A. There we find that the spectrum of perturbations contains two gapless phonon modes 
\begin{equation}\label{spectrum}
\omega_\pm= c_\pm(\alpha,\varepsilon) \,k\,,
\end{equation}
with the sound speeds bearing a nonlinear dependence on the strain parameters $\alpha,\varepsilon$. For the consistency and stability of a given $V(X,Z)$ around the background \eqref{finitedef} we require the absence of: $i)$ modes with negative kinetic energy, \textit{i.e.} ghosts; $ii)$ negative sound speeds squared, \textit{i.e.} gradient instability; $iii)$ superluminal propagation. In each case this leads to a certain value of maximal strain, $\varepsilon_{\text{max}}$, beyond which one of these consistency conditions is violated.
A typical stress-strain curve exhibiting this behavior, obtained for a given choice of $V(X,Z)$, is shown in Fig.~\ref{fig5}.

%For this and other time independent field configurations $\phi^I=\phi^I(x^i)$ the stress-energy tensor components are

%For the configurations \eqref{finitedef}, the stress-energy tensor components are
%\begin{align}\label{rho}
%&T^t_t\,\equiv\,\rho\,=\,V\,,\\\label{pressure}
%& T^x_x\,\equiv\,-\,p\,=\,V\,-\,X\,V_X\,-\,2\,Z\,V_Z\,,\\\label{txy}
%&T^x_{y}\,=\,2\,\partial_x \phi^I \partial_y \phi^I\,V_X\,,
%\end{align}
%where $V_X\equiv\partial V /\partial X,$ etc.

It is important to remark that our expressions for $\varepsilon_{\text{max}}$ derived in Appendix A should be interpreted as giving an \textit{upper bound} on the maximum strain that the material can support, since other effects not included here can enter before, thus lowering the actual maximum $\varepsilon$. For instance, one expects plastic/dissipative effects to enter at some point in real materials. However, this alters our analysis only for $\varepsilon_{\text{plastic}} < \varepsilon_{\text{max}}$, thus we still obtain an upper bound on  the maximum reversible deformability. %$\varepsilon_{\text{plastic}}$. 

It is interesting to consider the possibility that it really is
the $\varepsilon_{\text{max}}$ found here (or a value very close to it) that corresponds to the physical limitation to the material deformation.
%The fate of the  material past $\varepsilon_{\text{max}}$ is not entirely within the reach of the lowest order Lagrangian given in~\eqref{action}.
%If $\varepsilon_{\text{max}}$ is reached due to a gradient instability, the outcome also depends on the type of the tentative higher order stabilizing terms. In the case when they give rise to a positive $k^4$ correction to the phonon frequency squared, the instability will be slow and soft. 
%
In this case, the EFT gives partial information on {\it how} the material might `break'. 
As was mentioned earlier, there are two main options: that the breakdown is due to gradient instability or due to reaching superluminality.

In the case of gradient instability one expects that, like any instability, this is physically resolved by a transition to another ground state, most likely described by a different EFT. 
The specific nature of this transition remains hidden in the leading order low energy EFT presented in this work. 
For instance, whether the gradient  instability develops as a soft (slow) or hard (fast) process
depends on the nature of the next-to-leading order corrections to $V(X,Z)$. 
One may speculate that the hard case
corresponds to a breaking of the material and  the soft case
to the \textit{necking} phenomenon -- a decrease in the cross-sectional area of a material sample that is often seen under tensile stress. 
This would resemble the so-called `soft phonon' instability observed in some materials, see \cite{PhysRevB.76.064120,PhysRevLett.58.1035,PhysRevB.38.185,PhysRevLett.59.1329,PhysRevLett.78.4063,RevModPhys.46.83,PhysRevLett.91.135501,PhysRevB.89.184111}.
%
%See, \textit{e.g.}, \cite{PhysRevB.76.064120,PhysRevLett.58.1035,PhysRevB.38.185,PhysRevLett.59.1329,PhysRevLett.78.4063,RevModPhys.46.83,PhysRevLett.91.135501,PhysRevB.89.184111} 
%for some references on how the failure of certain solids proceeds via the so-called soft phonon instability.
%In that case, the onset of the instability does not yet correspond to failure, however it certainly marks a limit to the elastic behaviour.

Concerning superluminality, let us emphasize that in contrast to ghost and gradient instabilities the issue of superluminal propagation  relates to the possibility of a Lorentz invariant UV completion, not to the stability of propagation \cite{Adams:2006sv}. In order to apprehend the physical picture, it is instructive to recall a classic in field theory: the example given by  high spin fields where the problem of superluminality is known to arise~\cite{Velo:1970ur}. 
As discussed in \cite{Porrati:2009bs,Porrati:2011uu}, there are two ways to resolve the problem, which require to augment the EFT either by higher order operators or with additional light degrees of freedom. 
Any of the two resolutions makes it manifest that the naive EFT truncation (akin to the one that we are doing in  Eq.~\eqref{action}) breaks down. Moreover, it also gives an idea of {\it how} -- what that truncation might be missing.
In our case, this means that in the vicinity of violating the no-superluminality condition, corrections to the particular shape of $V(X,Z)$ that we consider must become important either by the presence of additional operators or light fields. 
The possibility that higher order operators (with more derivatives) can fix the superluminality problem while keeping the rest of the elastic response properties is nontrivial and we leave it for future research. On the other hand, the possibility that one needs to supplement the benchmark model with other light degrees of freedom seems quite reasonable 
-- after all in real world materials phonons do couple to many other modes. If this is the resolution, then the physical interpretation of the bound given by superluminality is that  $\varepsilon_{\text{max}}$ can be understood as an upper limit on when these light degrees of freedom have to be taken into account. %Again, $\varepsilon_{\text{max}}$ can be understood as an upper limit on when this can happen.

\section{Results in a scaling model}

For concreteness we shall focus on the simple potential
\begin{equation}
    V(X,Z)\,=\,\rho_{\text{eq}}\,\,X^A\,Z^{(B-A)/2}\,,\label{bench}
\end{equation}
where $\rho_{\text{eq}}$ is the dimensionful energy density set by the equilibrium configuration. 
The reason for choosing this form is that it realizes a power-law scaling like \eqref{power} at large deformations, $\varepsilon\gg1$. This behavior is observed in hyperelastic rubber-like materials, and there are many phenomenological models \cite{ZAMM:ZAMM19850650903,international2002atlas,doi:10.1063/1.1712836,23bdbaeed444470dbccb1db7302606d3,Ogden565,doi:10.5254/1.3538357,TF9444000059,1975JPhD....8.1285J}  that reduce to \eqref{bench} at large strains with various {\it strain exponents}~$\nu$. Here we are interested in characterizing how the stress-strain curves (and mainly the maximum stress and strain) depend on the parameters $A,B$. Let us also note that there are two special `corners' in parameter space: for $A=0$ the benchmark potential describes a perfect fluid \cite{Dubovsky:2005xd,Nicolis:2013lma}; for $A=1$, $B=1$ the model reduces to two free scalar fields.
%with only one %dynamical \OP{propagating} mode.

%\MB{This choice is inspired by the already existing  \textit{hyperleastic constitutive models} for rubber-like materials \cite{doi:10.1063/1.1712836,23bdbaeed444470dbccb1db7302606d3,Ogden565,doi:10.5254/1.3538357} which appear to fit within a very high accuracy \cite{doi:10.5254/1.3547969,doi:10.5254/1.3547847,Ogden2004} the experimental data \cite{TF9444000059,1975JPhD....8.1285J}. Nevertheless, exploiting the EFT power, we make a step forward defining the maximum strain and its correlation with the large strain scalings from first principles and not as just phenomenological inputs.}

We first find that the linear elastic moduli for the potential~\eqref{bench} take the simple form
\begin{equation}\label{KGlin}
{G}\,=\,\rho_{\text{eq}}\,2^A\,A\,,\quad
{K}\,=\,\rho_{\text{eq}}\,2^{A}\,B\,(B-1)\,.
%{K}\,=\,\rho_{\text{eq}}\,2^{1+A-2B}\,B\,(B-1)\,.
\end{equation}
They are both positive for $A>0$, $B>1$. Moreover the Poisson's ratio -- the negative ratio of transverse to axial strain -- for our models as \cite{Thorpe531} is readily obtained as:
\begin{equation}
\mathfrak{r}\,\equiv\,\frac{K-G}{K+G}\,=\,\frac{B(B-1)-A}{B(B-1)+A}\label{ff}\,.
\end{equation}
\begin{figure}[t]
    \centering
    \includegraphics[width=0.4\textwidth]{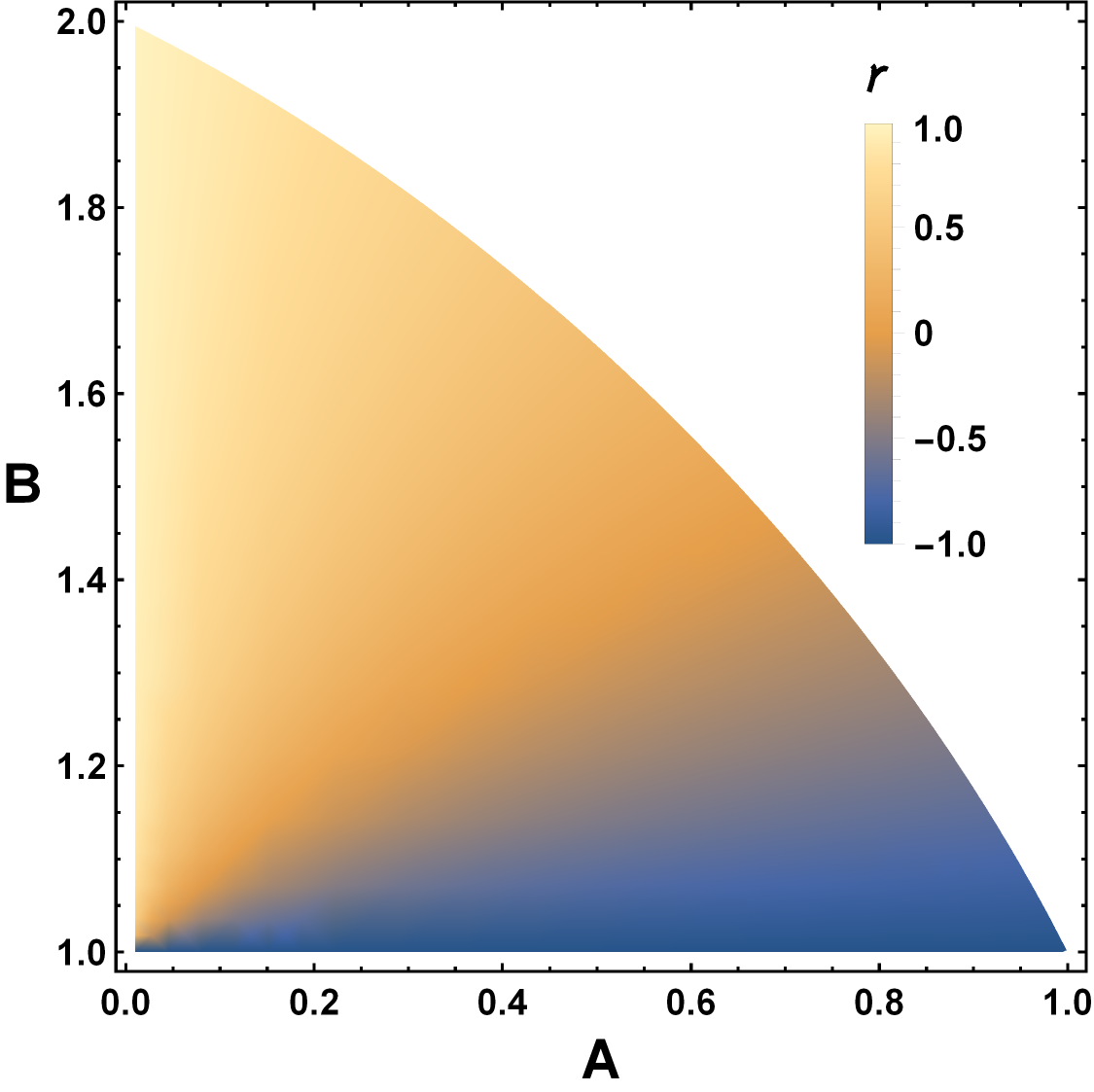}
   \caption{Poisson's ratio $\mathfrak{r}$ in the allowed parameter region given in Eq.~\eqref{full}.}
    \label{ratio}
\end{figure}
The result is shown in Fig.~\ref{ratio}. At large $B$ the ratio is close to its upper bound meaning that the models are close to perfect incompressible elastic materials. At small values of $B$ and large $A$ the ratio tends to its lower negative bound. A negative Poisson's ratio is typical of more exotic (the so-called {\it auxetic}) materials like some foams and metamaterials. Interestingly, the limit of free canonical scalars is in that regime. 
 Finally, for most of the models described, $-0.5<\mathfrak{r}<0.5$, as is common for steels and rigid polymers.

\begin{figure}
    \centering
    \includegraphics[width=0.4\textwidth]{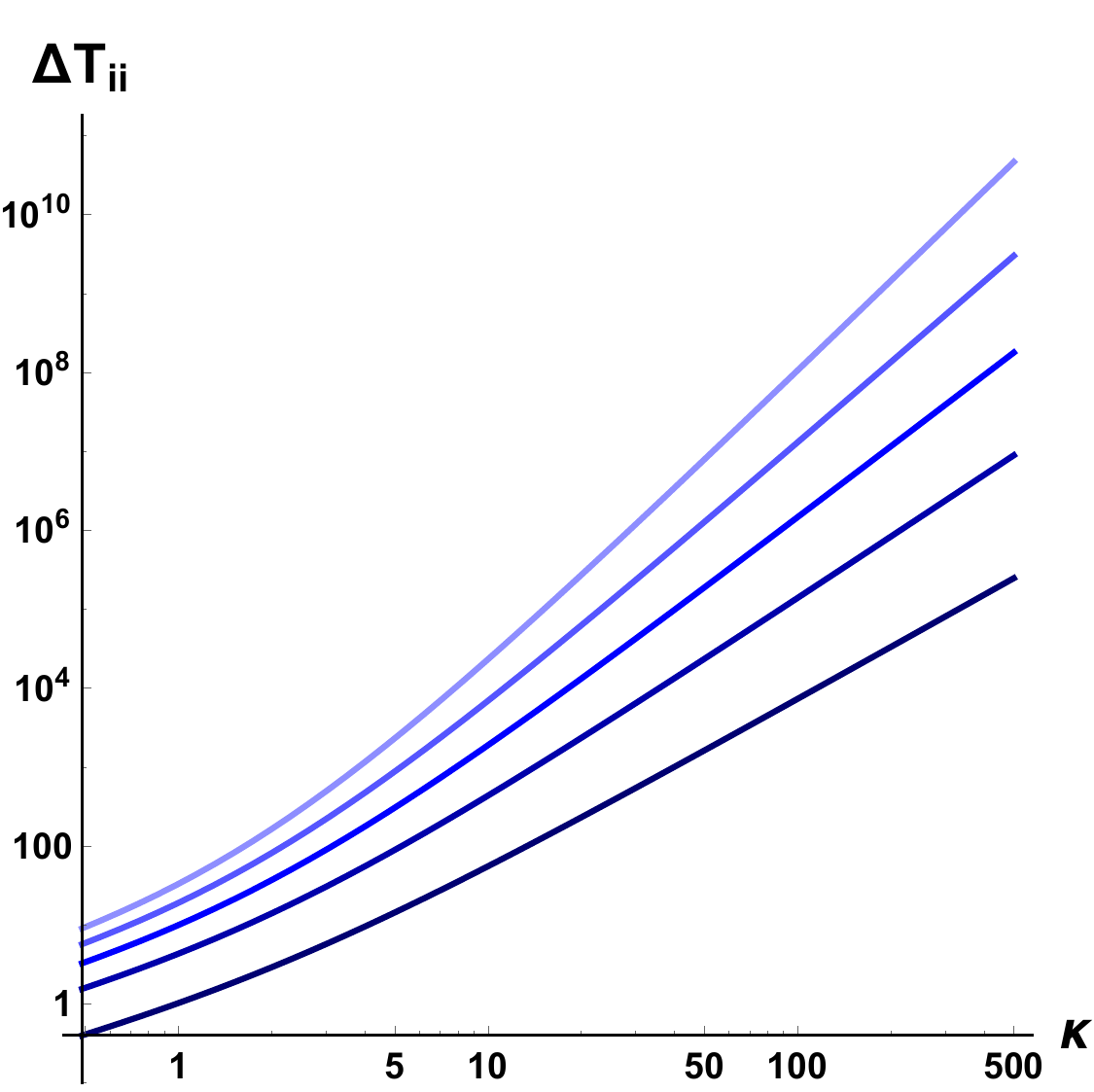}
    \caption{The nonlinear bulk stress-strain curve for the benchmark model and the parameter values $A=0.5$ and $B=1.1,1.3,1.5,1.7,1.9$. The large strain scaling is set by $\Delta T_{ii} \propto \kappa^{2B}$.}
    \label{fig6}
\end{figure}

For the full nonlinear response to pure shear, Eq.~\eqref{sigma} gives:
\begin{equation}\label{sigma2}
   \sigma(\varepsilon)\,=\, \rho_{\text{eq}}\,A\, \varepsilon\,  \sqrt{\varepsilon ^2+4} \left(\varepsilon ^2+2\right)^{A-1}\,.
\end{equation}
This is shown in Fig.~\ref{fig5} for various values of $A$ and $B$. %along with the breaking points arising as a consequence of the onset of an instability. 
Notably, the stress-strain curves obtained from the benchmark models mimic a large variety of materials including fibers, glasses and elastomers \cite{international2002atlas}. More precisely, Eq.~\eqref{sigma2} describes Neo-Hookean systems which follow Hooke's law at small strain but exhibit non-linear power-law scalings at large deformations \cite{Ogden2004}. Similarly, the nonlinear response to a pure \textit{compression}, that we define as $\kappa\equiv\alpha-1$, reads
\begin{align}
    \Delta T_{ii}(\kappa)=\rho_{\text{eq}}\,2^{A+1}\,(B-1) 
   \left[(\kappa+1)^{2B}-1\right]\,.
\end{align}
We show the full nonlinear response to pure bulk deformation for various values of $B$ in Fig.~\ref{fig6}.
As per construction, at large strains, $\varepsilon,\kappa \gg 1$, the nonlinear stresses display power-law scalings of the form: 
\begin{equation}\label{nonlin}
    \sigma(\varepsilon)\,\sim\,A\,\varepsilon^{2A}\,,\quad \Delta T_{ii}(\kappa)\,\sim\,(B-1)\,\kappa^{2B}\,,
\end{equation}
from where we read off the shear and bulk strain exponents as: $\nu_{\text{shear}}=2A$ and $\nu_{\text{bulk}}=2B$.
Note that, as can be seen from Eq.~\eqref{KGlin}, $A$ and $B$ also control the linear shear and bulk moduli. 

%This is however a non-generic feature arising due to the simple power-law dependence on $X$ and $Z$ of the potential $V(X,Z)$. In particular, the direct correlation between the powers and the moduli can be broken by adding new terms to the potential \eqref{bench}.  For the sake of simplicity we shall not do so here. Keeping this in mind, though, we shall stick to the physical meaning of $A$ and $B$ as the exponents  in Eq.~\eqref{nonlin}, because they characterize the large strain regime.

%constrains the shape of the potential.
%In addition to \OP{the stability conditions above}
%demanding the absence of ghosts, gradient instabilities and superluminal propagation 
%we shall also require positivity of the elastic moduli $K$ %and $G$. 

Combining the requirements of the absence of ghosts, gradient instabilities and superluminal propagation %\footnote{Let us remark that the issue of superluminal propagation is different from the others in the sense that it relates to the possibility of a Lorentz invariant UV completion, not to the stability of propagation \cite{Adams:2006sv}. }
%\textcolor{red}{For instance, one of the standard field theory examples where the problem of superluminality arises is in the context of higher spin fields \cite{Velo:1970ur}. There the resolution comes from introducing either higher order non-minimal couplings between the fields or new light degrees of freedom \cite{Porrati:2009bs,Porrati:2011uu}. In either case, it corresponds to providing some sort of UV completion that  renders the speed of propagation subluminal.  While finding such UV completion is certainly beyond the point that we wish to make in the present work, the appearance of superluminality is nevertheless a clear indication that we are exiting the regime in which our EFT (written only for the low energy phonons) is the correct description of the material.}
 with the  positivity of the elastic moduli, $K$ and $G$, constrains the allowed range of parameters. In the simple case of linear deformations we obtain the following allowed region for the exponents $A,B$:
\begin{equation}
    0\leq A\leq 1\quad \text{and} \quad  1\leq B\leq \sqrt{1-A}+1 \label{full}\,.
\end{equation}
%\begin{itemize}
%    \item $A \geq 0$ comes from the requirement of no gradient instabilities which coincides with the positivity of the shear elastic modulus
%    \item $B\geq 1$ comes from the positivity of the linear bulk modulus
%    \item $A \leq 2B-B^2$ comes from the requirement of superluminality at zero strain.
%\end{itemize}

The analysis can be extended to finite strain and as mentioned above leads us to another important result: the existence of a maximum strain 
$\varepsilon_{\text{max}}$ that can be supported by the system before the onset of one of the aforementioned pathologies. How $\varepsilon_{\text{max}}$ depends on the strain exponents is shown in  Fig.\ref{fig1}; the exact analytic expressions can be found in the Appendix.\footnote{Let us remark that, as can be inferred from Eq.~\eqref{sigma2}, the power-law scaling can really be reached only  for $\varepsilon \gtrsim 2$. Therefore, the limits shown in Fig.~\ref{fig1} can only be extended to a material following \eqref{nonlin} at large strains in the bluish part of the diagram.}
We must emphasize that the $\varepsilon_{\text{max}}$ obtained in this way is not meant to be the actual maximum deformation that a material with the aforementioned scaling properties can withstand, but rather an upper bound on 
it. %\footnote{
%For instance, one expects plastic/dissipative effects to enter at some point in real materials. However, this alters our analysis only for $\varepsilon_{\text{plastic}} < \varepsilon_{\text{max}}$, thus we obtain an upper bound on  $\varepsilon_{\text{plastic}}$. 
%%
%Another implicit assumption in our analysis is that the EFT remains a valid truncation at large $\varepsilon$. 
%This could be violated if there are any other light degrees of freedom that couple to strain, or which become light at some finite strain. Again, $\varepsilon_{\text{max}}$ can be understood as an upper limit on when this can happen.}
%
Still, this already provides quite a lot of information. For instance, in the large (yellowish) area of Fig.~\ref{fig1} where  
$\varepsilon_{\text{max}}$ only reaches values of~$\sim1$, one can already discard the existence of very elastic materials that exhibit scaling as in \eqref{power} with those scaling exponents.

We note that the regions in the $A-B$ parameter space where large strains can be supported are near the special points  
$A=1,B=1$ (free scalars)
or $A=0$ (\textit{fluid} limit).
Therefore, for the model \eqref{bench}, we expect the real-world (non-relativistic) solids to lie near the $A=0$ axis. 
%\textcolor{blue}{This shows clearly that a minimally coupled scalar cannot describe a real material. On the other hand it is known that a perfect fluid does not respond to static shear deformations and thus naturally can support infinite strain.} 
In this limit,  the maximum strain is set from the absence of gradient instability for almost all values of $B$. 

Intriguingly, for $A\ll1$ a number of `universal' correlations appear.
First,  we find a universal scaling of  the maximum strain 
   \begin{equation}\label{emaxfluid}
    \varepsilon_{\text{max}}\simeq\sqrt{2}\left(\frac{B-1}{A}\right)^{1/4}\,.
\end{equation}
Inserting this in the expression \eqref{sigma} for the nonlinear shear stress we further obtain
\begin{equation}
    \sigma_{\text{max}}\equiv\sigma(\varepsilon_{\text{max}})=\rho_{\text{eq}}\,A\,.
\end{equation}
This shows a linear dependence  of the maximal stress supported by a material %(sometimes referred to as {\it ultimate tensile strength} UTS)
on the strain exponent $A$, which in our simple model controls also the linear elastic modulus.  
%as is clearly seen also from the Fig.~\ref{fig9} in the limit $A\to 0$. 
Similar linear correlations are  observed experimentally in various materials \cite{MEMI27,article,ZHANG201162,doi:10.1063/1.3544202,doi:10.1179/174328405X29302}.
Additionally, we also find a clear relation between the hardness and the maximum strain, %(\textit{strength}), 
$\sigma_{\text{max}}\sim\varepsilon_{\text{max}}^{-4}$. 
Let us emphasize, however, that whether the correlations that we find can be extrapolated to real world materials strongly depends on $i)$ whether their stress-energy function $V$ behaves as a power law at large strain, and $ii)$~whether they can support large deformations.

\begin{figure}[t]
    \centering
    \includegraphics[width=0.4\textwidth]{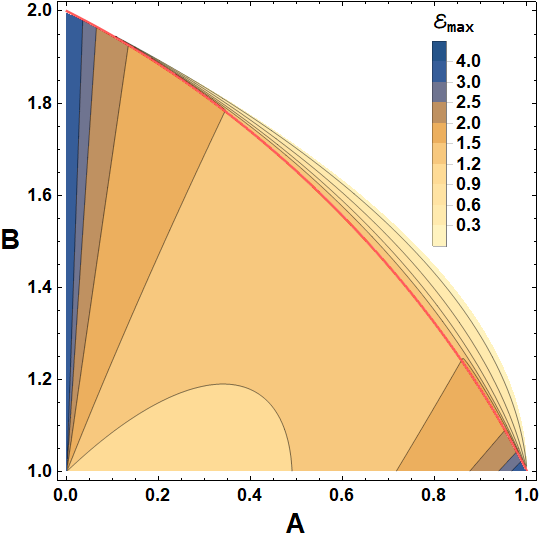}
    \caption{The allowed parameter region \eqref{full} for the benchmark model \eqref{bench}. The left, bottom and right edges are respectively given by: gradient instability, positivity of the bulk modulus, superluminality.  The red line separates the region where the maximum strain is due to the gradient instability (left) and the region where it is due to superluminality (right). Large strains (and therefore the power-law behaviour \eqref{power}) are realized in the bluish area.}
    \label{fig1}
\end{figure}

%\begin{figure}[th]
 %   \centering
  %  \includegraphics[width=0.30\textwidth]{stressmax2.png}
   % \caption{Maximum stress that can be withstand by a material before the onset of the gradient instability as a function of $A$ for $B=1.01,1.5,1.9$ and $\rho_{\text{eq}}=1$ in the logarithmic scale.
    %Maximum strain as a function of $A$ for $B=1.01,1.5,1.9$. The discontinuities are given by the saturation of the inequality \eqref{sep}. The full line indicates that the maximum strain arises due to a gradient instability; the dashed lines indicate the onset of a regime where superluminality is the main source of instability.
 %   }
  %  \label{fig9}
%\end{figure}

Finally, let us note that within the benchmark model \eqref{bench} there are no constraints on the bulk strain $\kappa$ arising from the consistency and stability requirements. This is a consequence of \eqref{bench} being a monomial. For more general choices, additional bounds can arise. Let us also mention that for $B\in(0,1)$ it is possible to achieve a negative bulk modulus, $K<0$, in a way that is perfectly consistent from the EFT perspective. In particular, as long as $K>-G$ the stability constraint $c_+^2>0$ is still satisfied. This has also been studied in four dimensions \cite{doi:10.1177/0021998305051112,doi:10.1002/pssb.200777708} and observed experimentally \cite{doi:10.1080/09500830600957340}.

\section{Non-relativistic solids}\label{nonrelativistic}
The benchmark model Eq.~\eqref{bench} considered above has been useful to exhibit the constraining power of the EFT methods, however it has one disadvantage: the region in parameter space giving small sound speeds as in real-world elastic materials is very small.  To be more specific, the typical sound speeds are at most of order $\sim 10^{-4}$ in the units of the speed of light. In the parameter space $A, B$, this corresponds to the corner where both $A$ and $B-1$ are of order $10^{-8}$, or less. 
The problem with this is that in the benchmark model \eqref{bench} $A$ and $B$  also control the exponents in the stress-strain relation at large strain, $\sigma \sim \varepsilon^{\nu}$ with $\nu_{\text{shear}}=2A$ for pure shear and $\nu_{\text{bulk}}=2B$ for pure bulk deformations respectively. 
It follows that the benchmark models can only cover realistic materials with very specific exponents, basically $\nu_{\text{shear}}\sim 10^{-8}$ and $\nu_{\text{bulk}}\simeq2$. 
Clearly, there has to be a way around this limitation because elastic materials with more generic values for $\nu_{\text{bulk/shear}}$ do exist and one expects that a similar EFT construction should describe them. The obvious guess is that the benchmark choice Eq.~\eqref{bench} is too restrictive. In this section we show how to deform the model in order to have small speeds of sound while keeping large deformability and generic exponents. 

Fortunately, there is a well motivated and unique way to ensure that the sound speeds become as small as needed while preserving the stress-strain relations untouched. 
This is achieved by adding an extra term to the potential $\delta V \propto \sqrt{Z}$ with a large coefficient in front. 
This term is special for many reasons. Physically, it is proportional to the mass density of the material \cite{Leutwyler:1996er}. This immediately explains why the coefficient in front of it must be large in the non-relativistic materials. 
The mass density contributes to the Lagrangian (an energy density) weighted by $c^2$  \cite{Leutwyler:1996er} and is much larger than the typical stresses in solids. 
Related to this, in the  fluctuations around any background, this  term only produces temporal kinetic terms,
as can be easily seen in equations \eqref{pert1}-\eqref{pert2} in Appendix A, noting that this term satisfies $\delta V_Z+2Z\delta V_{ZZ}=0$. 
Therefore, this new term only contributes to the denominators in the formulas for the speeds, and so enhancing it  decreases the speeds. 

Moreover, an important feature of this term is that it does not affect neither the bulk stress $T_{ii}$ nor the shear stress $T_{ij}$, so it doesn't alter the stress-strain relations (this is clear from Eqs.~\eqref{pressure} and~\eqref{sigma}). This term only appears in the energy density $T_{00}$, as it must be, since it only accounts for the inertial mass and thus it contributes like `dust' (pressureless fluid). This is crucial to retain the predictive/constraining power of the EFT framework, because in order to go to the non-relativistic regime it suffices to add one single parameter in the full nonlinear Lagrangian.

For these reasons, it suffices to switch to the following model,
\begin{equation}\label{bench2}
V(X,Z) \,=\, \rho_0\, \left(\sqrt{Z}\,+\, v^2 \left(\frac{X}{2}\right)^A\,Z^{\frac{B-A}{2}}\right)~,
\end{equation}
with $v$ a small parameter (which is a measure of the typical speeds in the units of the speed of light). This guarantees that the material is non-relativistic while keeping the nonlinear static elastic response the same as in the benchmark model \eqref{bench}.

\begin{figure}[htpb]
    \centering
    \includegraphics[width=0.4\textwidth]{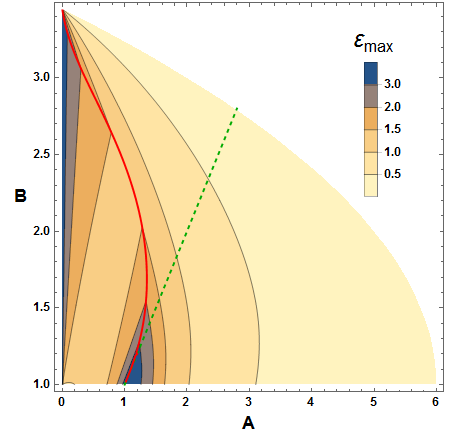}
    \caption{Expanded parameter space for $v^2 = 0.2$. The red line splits the regions where the limit on the maximal strain comes from superluminality (on the right) and from gradient instability (on the left). The green dashed line is $A=B$. In the region $A\geq B$ the maximum strain is only dictated by subluminality.}
    \label{emaxcase2}
\end{figure}
\begin{figure}[htpb]
    \centering
    \includegraphics[width=0.4\textwidth]{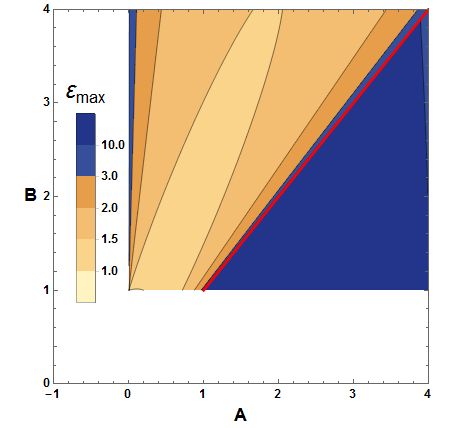}
    \caption{Expanded parameter space for $v^2 = 10^{-8}$. The subluminal constraint in $A<B$ is now located at larger values of $A$ and $B$. In the region $A\geq B$ the maximum strain is only dictated by subluminality.}
    \label{emaxcase1}
\end{figure}

The discussion about the stability and consistency of this model is also mentioned in the Appendix. In summary, we find that for $v\ll 1$ there are {\it two} new regions in the $A-B$ parameter space that allow $i)$ small velocities and $ii)$ $\varepsilon \gg 1$ (\textit{i.e.} a very elastic material), as can be seen in Figs.~\ref{emaxcase2}, \ref{emaxcase1}.
The first region is close to the line $A = B$ but with ($A<B$). The other one extends for  $A>B$ relatively close to $A=1$. 
These two regions are conceptually on a different level from the EFT standpoint because the constraint on $\varepsilon$ arises from gradient instability or superluminality in either case. The separatrix between the two regions is given by the red line in the plots. For $v\ll1$ this line is very close to the line $A=B$ at small $A, B$. Importantly, both regions contain sizeable values for $A, B$.
The first conclusion, then, is that, indeed, adding a large mass-density term $\sqrt Z$ to the Lagrangian opens up the possibility to model non-relativistic materials with sizeable shear and bulk exponents, $\nu_{\text{shear}}=2A$ and $\nu_{\text{bulk}}=2B$.

In the new region with $A<B$, the subluminality condition does not play any role (for $v\ll1$ and moderate values of $A,B$), so these are reasonable candidate EFTs to model realistic materials. In this region, the EFT  again `predicts' that the maximum strain  $\varepsilon_{\text{max}}$ and the bulk/shear exponents are related in a simple way. 
One can see 
%from \eqref{emax} 
that $\varepsilon_{\text{max}}$ scales with the exponents as
\begin{equation}
\varepsilon_{\text{max}} \sim \sqrt2 \,\left(\frac{A}{B-A} \right)^{1/4}\,,\qquad A<B\,.
\label{emaxgrad}
\end{equation}
Interestingly enough, even though this differs from Eq.~\eqref{emaxfluid} (valid for the benchmark model \eqref{bench} at $A\ll1$) we still have some relation $\varepsilon_{\text{max}} (A,B)$.

The second new region (for $A>B$) instead is only constrained by the subluminality condition, and so the bounds are less powerful. Specifically,  for $v \ll 1$ we find 
\begin{equation}
\varepsilon_{\text{max}}^2 \sim 2\,\frac{1}{v^{2/A}}\left(A\,(A+1)\,+\,B\,(B-3)\right)^{-1/A}.
\end{equation}
Notice that the maximum strain scales as $\varepsilon_{\text{max}}^{\nu_S}\sim 1/v^2$, where $\nu_S\equiv 2A$. %The fact that $\varepsilon_{\text{max}}$ grows this way with $v^2$ is because 
This scaling can be understood because for large shear deformation the phonon speed $c_+$ grows as $c_+^2 \sim v^2 \varepsilon^{2A}$. Since the constraint obtained within the EFT is really only an upper limit on the strain, one obtains only a very large upper bound -- a very loose bound.\\
\begin{figure}[htpb]
    \centering
    \includegraphics[width=0.4\textwidth]{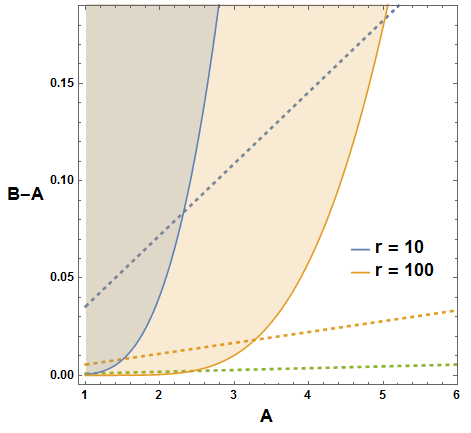}
    \caption{Speed ratio constraints for $r = 10, 10^2$. Dashed lines show where $\varepsilon_{\text{max}}= 3, 5, 8$ (Blue, Orange, Green).}
    \label{speedratio}
\end{figure}

As is clear from Figs.~\ref{emaxcase2}, \ref{emaxcase1}, the two regions actually touch each other, therefore at some point one of the speeds must increase also in the $A<B$ region. Since in this region $\varepsilon_{\text{max}}$ comes from the  gradient instability of one of the two modes, \textit{i.e.} by setting one of the speeds $c_-(\varepsilon_{\text{max}})=0$, a good notion of how non-relativistic the material is at the `breaking' point is given by the other phonon speed, \textit{i.e.} $c_+(\varepsilon_{\text{max}})$. At $\varepsilon=0$ all the speeds are granted to be of order $v$, while, by definition, $c_+(\varepsilon_{\text{max}})=1$  on the separatrix between the two types of the new regions. However in truly non-relativistic materials one does not expect $c_+(\varepsilon_{\text{max}})$ to raise to  such large values. 
Therefore, in order to be more realistic, we can also impose that the speeds do not vary much from their values at equilibrium ($\varepsilon=0$) to the breaking point. To this end, we introduce the ratio
\begin{equation}\label{r}
r \, \equiv \,\frac{c_{+}(\varepsilon_{\text{max}})}{c_{+}(\varepsilon = 0)} 
\end{equation}
and demand that in the region $A<B$ it is allowed to grow at most by a factor $10-10^2$. In the limit $v \ll 1$, $r$ is only a function of  $A$ and $B$, and in the region $B\sim A$ we find, using \eqref{emaxgrad},
\begin{equation}
r \sim \left(B-A\right)^{-A/4}.
\end{equation}
This new constraint is shown in Fig. \ref{speedratio}. This figure shows that there is indeed an overlap between the regions corresponding to large deformation ($\varepsilon_{\text{max}}$ significantly greater than 1), moderate $r$ and large exponents. The size of this region in parameter space depends on the criteria for $r$ and $\varepsilon_{\text{max}}$, but one can say that it extends to next to the $A=B$ line within a few percent. 
In this region, non-trivial relations such as \eqref{r} or \eqref{emaxgrad} should hold. 

%As a final remark, let us summarize the most basic `predictions' from the EFT method on realistic and very elastic materials that display a power-law stress strain curve both for shear and bulk deformations.
%First, as is clear from Fig.~\ref{emaxcase1}, there are should be no such materials with shear exponent in the window 
%$10^{-8} < \nu_{\text{shear}} < 2$, where the lower bound is just an order of magnitude estimate. This 

As a final remark, let us emphasize the most basic property of this new region: it is close to $A = B$. 
In other words, this corresponds to very elastic realistic materials that display a power-law stress strain curve both for shear and bulk deformations, with nearly equal  bulk and shear exponents, $\nu_{\text{shear}}\simeq \nu_{\text{bulk}}$.

\section{Discussion}
In conclusion, let us highlight that EFT methods for solid materials allow to extract 
%describing the low energy dynamics of the lightest degrees of freedom in solids -- the phonons -- contain 
%a material dependent 
non-trivial information and bounds on their nonlinear elastic response.
%One first main point is that even though the symmetry breaking pattern present in solids only restricts that the effective Lagrangian is specified by a basically free function $V(X,Z)$, for every given material,  the resulting  EFT is fully predictive because the full function $V(X,Z)$ can be directly extracted from the measured stress-strain diagrams. 
%
%One our main findings presented here clarify in what sense an EFT for the Goldstones of spontaneously broken spacetime symmetries, written in terms of a function $V(X,Z)$ can still be predictive. The answer to this question is that the full function $V(X,Z)$ can be in principle measured directly from the stress-strain diagrams and from the phonons sound speed dependence on the applied strain. 
%The main motivation of our work was to clarify how an EFT that involves an arbitrary function $V(X,Z)$ can still be predictive. The answer to this question is that the full function $V(X,Z)$ can be in principle measured directly from the stress-strain diagrams.
%
%Our main point outlined in this Letter is that 
%
%Therefore, treating the classic elasticity theory as a standard EFT provides a useful tool to describe both linear and nonlinear effects and the various correlations among them. 
%
The list of observables that are fixed (to the leading order in the EFT) once the strain-energy function $V(X,Z)$ is known includes: %, for instance, 
all the $n$-point phonon correlation functions, the phonon-phonon self-interactions and, most remarkably, how these depend on the applied stresses -- the first example of this being the acoustoelastic effect. %(how the phonons speeds depend on applied strains). 
%This set of correlations is, of course, most directly ...
The correlations obtained in this way are most directly relevant for materials that admit large deformations and where dissipative effects are unimportant.\footnote{For recent EFT-like efforts to include dissipation in fluids and viscoelastic materials, see  \cite{Grozdanov:2013dba,Haehl:2014zda,Haehl:2017zac,Haehl:2018lcu,Glorioso:2018wxw,Crossley:2015evo,deBoer:2015ija,2016RPPh...79a6502T,2017PhRvE..96f2134T,Beekman:2017brx,Beekman:2016szb,Armas:2018ibg,Jensen:2017kzi,Jensen:2018hse}.}

%While some of these results were partially appreciated in the literature, we find that the EFT language is useful and can give nontrivial fruits.  
As a specific application we have studied how the maximal strain supported by a given material is constrained by the consistency of the EFT. % effective field theory. 
Focussing on the class of materials with power-law stress-strain relations at large strains, $\sigma\sim\varepsilon^\nu$,
%For such power-law behaviour 
we find several universal relations between intrinsically nonlinear response parameters, such as the maximum stress and the strain exponent.

%the maximum stress and the strain hardening exponent $\nu$, which holds when both of them are small. 
%the exponent $\nu=2A$, where the latter is associated with the stiffness of the material. 
%
%Let us however point out that our findings apply only under the condition that the EFT is a suitable description of the material, which implicitly neglects any dissipative effects. 
%
%
%Therefore our results can be viewed to set where the plastic/dissipative regime should enter.\\
%Finally it would be valuable to explore the fracton-elasticity duality \cite{2017arXiv171111044P} in our language and construct the nonlinear EFT for fractons.

An interesting  case is represented by the {\it  conformal solids} limit, realized by potentials of the form $V(X,Z)\,=\,X^{3/2}\,{F}\left(X/Z^{1/2}\right)$, which preserve scale invariance and imply $T^\mu_\mu=0$ \cite{Esposito:2017qpj} (see also \cite{Alberte:2017cch,Baggioli:2018bfa}).
%
%Let us mention here that in the conformal 
In this case, the bulk modulus is directly proportional to the energy density  $K=3/4\,\rho$, as observed in earlier holographic models \cite{Baggioli:2018bfa}.
Concerning the strain exponents, scale invariance fixes  $\nu_{\text{bulk}}=3$ and bounds   $\nu_{\text{shear}}\leq 3/2$.
%which are in good agreement with what is observed in the  holographic realizations of critical materials \cite{holversion}.
Let us emphasize, however, that the notion of a {\it conformal solid}, understood as an EFT with a Lagrangian of the above form, should be distinguished from a system whose low energy dynamics is controlled by a strongly coupled infrared fixed point. In that case, the standard EFT methods are not granted to apply. A study of the nonlinear elasticity for that case using holographic techniques is deferred to a separate work \cite{holversion}. 

We have also shown how to extend the analysis to non-relativistic materials, with realistically small sound speeds. Our main conclusion -- that the EFT method  provides nontrivial relations between nonlinear response parameters -- remains true also in this regime. Moreover, let us make a remark about the region close to $A=B$ of these non-relativistic solids. In this region, the EFT method is the most informative, so it is worth trying to compare its predictions to data. 
A proper analysis of the experimental data on real world elastomers is well beyond the scope of this work, but we would like to make one comment. 
%(after numerous phenomenological models to fit the nonlinear elastic response of rubbery materials) 
It is known \cite{enlighten70333} that a very successful way to fit the nonlinear response of some rubbers
consists of writing $V(X,Z)$ as a sum of a few powers of the matrix $X^{IJ} = \partial_\mu \phi^I \partial^\mu\phi^J$ as
$V=\Sigma_n \mu_n \text{Tr}[(X^{IJ})^{p_n}]$ with some constants $\mu_n, p_n$. 
At large deformations, these models are dominated by the term with the highest power, call it $p$. 
It is easy to see that taking $V=\text{Tr}[(X^{IJ})^{p}]$ does not  strictly coincide with our benchmark models for any $A,\,B$, 
however, it does lead to very similar response at large strains as our benchmark model with $A= B =p$ (for instance in the stress-strain relations). This is encouraging because it would suggest that  the models in the region near $A=B$ could correspond to these rubbers. It would be interesting to see whether \eqref{r} or \eqref{emaxgrad} hold for them. We leave these questions for the future.

%Let us emphasize that there are two very different notions of {\it conformal solids}. The first one refers to solid materials that can be described with EFT methods, the potential \eqref{confpot} being a particular example. The existence of such an EFT description implicitly assumes a clear scale separation allowing for a low energy (or derivative) expansion. This means that the material possesses a scale which in the EFT language takes the role of the cutoff scale, $\Lambda$, and which in the context of mechanical response can be identified as the materials energy density scale. Furthermore, one can assume that up to energies of order $\Lambda$ there are no other weakly coupled degrees of freedom than the phonons. 
%The second (perhaps more realistic) notion applies when the low energy dynamics is controlled by a strongly coupled infrared fixed point. In that case, there is a continuum of low energy modes with no scale separation; therefore the standard EFT methods are not applicable. This case can, however, be addressed using holographic techniques and will be the subject of a separate work \cite{holversion}. 

Furthermore, it would be desirable to introduce dissipative and thermal effects within the EFT picture of condensed matter systems \cite{Endlich:2012vt,2017PhRvE..96f2134T}. In this regard the holographic description could provide a valuable supplementary insight \cite{Baggioli:2014roa,Baggioli:2018bfa,Alberte:2016xja,Alberte:2017cch,Andrade:2017cnc,Baggioli:2018bfa,holversion,Alberte:2015isw,Grozdanov:2018ewh}. %which can be even extended beyond the conformal limit \cite{Baggioli:2018bfa}. 
We hope to return to some of these points eventually.
\section*{Acknowledgements}
We thank Alex Buchel, Carlos Hoyos, Karl Landsteiner, Mikael Normann, Giuliano Panico, Napat Poovuttikul, 
Kostya Trachenko and Alessio Zaccone for useful
discussions and comments about this work and the topics
considered.
MB is supported in part by the Advanced ERC grant
SM-grav, No 669288.
VCC and OP acknowledge support by the Spanish Ministry
MEC under grant FPA2014-55613-P, FPA2017-88915-P and the Severo
Ochoa excellence program of MINECO (grant SO-2012-
0234, SEV-2016- 0588), as well as by the Generalitat
de Catalunya under grant 2014-SGR-1450.\\
MB would like to thank Iceland University and Queen Mary University for the warm hospitality during the completion of this work and Marianna Siouti for the unconditional support.

\appendix
\section{Fluctuations and consistency}\label{app:sound}

In order to study the stability of perturbations around the strained background configuration we expand the scalar fields as $\phi^I=\phi_{\text{str}}^I+\pi^I$. To identify the propagating degrees of freedom we perform the decomposition into longitudinal and transverse fluctuations by splitting $\pi^I = \pi^I_L+\pi^I_T$, with $\pi_{L/T}$ satisfying:
\begin{equation}
O^I_K\partial_I\pi^K_L = 0\,,\quad \varepsilon^{IJ}O_I^K\partial_K\pi^T_J=0\,.
\end{equation}
This gives two dynamical scalar modes that can be defined through:
\begin{equation}\label{modes}
    \pi^I_L = O^{IK}\partial_K\pi_L\,,\quad \pi^I_T = \varepsilon^{IJ}O_J^K\partial_K\pi^T\,.
\end{equation}
Constraining the spatial dependence to $\pi_{L/T}=\pi_{L/T}(t,x)$ and redefining $\pi_{L/T}\to\pi_{L/T}/\sqrt{-\partial_x^2}$ we obtain the following quadratic action for the fluctuations
\begin{align}
\delta S_2 = \int d^3x\,\Big[&N_T\dot\pi_T^2+N_L\dot\pi_L^2+2N_{TL}\dot\pi_T\dot\pi_L-c_T^2(\partial_x\pi_T)^2\nonumber\\&-c_L^2(\partial_x\pi_L)^2-2c^2_{TL}\partial_x\pi_T\partial_x\pi_L\Big]\,,\label{action2}
\end{align}
where the parameters $N_T,N_L,N_{TL}$ and $c_T^2,c_L^2,c_{TL}^2$ depend on both the shear and bulk strains, {\it i.e.}, they are functions of $\varepsilon$ and $\alpha$. The explicit expressions in terms of the derivatives of the function $V(X,Z)$, defined in Eq.~\eqref{bench}, are found to be:
\begin{align}
&N_T = \frac{1}{2}\left((X^2-2Z)V_Z+{X}V_X\right)\,,\label{pert1}\\
&N_L = Z \,V_Z+\frac{X}{2}V_X\,,\\
&N_{TL}=\frac{1}{2}\sqrt{Z(X^2-4Z)}V_Z\,,\\
&c_L^2=Z\left(V_Z+2ZV_{ZZ}\right)\\
&+\frac{1}{2}X\left(V_X+4ZV_{XZ}+XV_{XX}\right)\,,\nonumber\\
&c_T^2=\frac{1}{4}\left((X^2-4Z)(V_Z+2ZV_{ZZ})+2XV_X\right)\\
&c_{TL}^2 = \frac{1}{2}\sqrt{Z(X^2-4Z)}(V_Z+2ZV_{ZZ} +XV_{XZ})\,,\label{pert2}
\end{align}
with all the quantities evaluated on the scalar field background solution $\phi_{\text{str}}^I$.

Let us emphasize that for a non-diagonal  matrix $O^I_J$, the transverse and longitudinal modes remain mixed both with respect to time and spatial derivatives. In order to study the stability of fluctuations we therefore first introduce the kinetic matrix as
\begin{equation}
\mathcal{N}\,=\,\begin{pmatrix} 
N_T & N_{TL} \\
N_{TL} & N_L
\end{pmatrix}\,.
\end{equation}
The absence of ghost-like excitations then requires that the eigenvalues of the kinetic matrix, $\lambda_\pm$, are positive. This gives the first condition for stable propagation of the modes: $\lambda_{\pm}\,>\,0$.

It is straightforward to determine the true dynamical modes described by the action \eqref{action2} by working at the level of the equations of motion of the mixed fields $\pi_{L/T}$. After Fourier transforming as $\pi_{L/T}=a_{L/T}\,\mathrm{e}^{i\omega t-ikx}$ we can solve for the spectrum of perturbations to obtain
\begin{equation}\label{spectrum2}
\omega^2_\pm= c^2_\pm (\alpha,\varepsilon)\, k^2\,.
\end{equation}
The other conditions for consistency that we are going to impose are thus:
\begin{itemize}
    \item $c_{\pm}^2\,\geq\,0$, \textit{i.e.} the absence of gradient instabilities;
    \item $c_{\pm}^2\,\leq\,1$, \textit{i.e.} the absence of superluminal modes.
\end{itemize}

The exact expressions of the kinetic eigenvalues can be put in the form
\begin{equation}
    \lambda_{\pm} =\frac{c}{2}\left[1\pm\sqrt{1-\frac{4d}{c^2}}\right]\,,
\end{equation}
with 
\begin{align}
    c&=N_L+N_T\,,\\
    d&=N_TN_L-N^2_{TL}=\det\mathcal N\,.
\end{align}
Similarly the sound speeds can be expressed as
\begin{equation}\label{spectrum3}
c^2_\pm =\frac{a}{2d}\left[1\pm\sqrt{1-\frac{4bd}{a^2}}\right]\,
\end{equation}
with 
\begin{align}
&a = c_T^2N_L+c_L^2N_T-2c_{TL}^2N_{TL}\,,\\
&b = c_T^2c_L^2-c_{TL}^4\,.
\end{align}
Let us point out that evaluating the sound speeds $c_\pm$ at $\alpha = 1$ and $\varepsilon=0$ we find that the result coincides with the standard relationships obeyed by the transverse and longitudinal phonons of the equilibrium state $\phi^I_{\text{eq}} = x^I$: 
\begin{equation}
c_T\,=\,\sqrt{\frac{{G}}{\rho+p}}\,,\qquad c_L\,=\,\sqrt{\frac{K+G}{\rho+p}}\,,\label{speed0}
\end{equation}
where $\rho$ and $p$ are the equilibrium energy density and pressure, as in Eq.~\eqref{rho} and \eqref{pressure}. The $K$ and $G$ refer to the linearized bulk and shear moduli, defined in Eq.~\eqref{Klin} and \eqref{Glin}.

The conditions necessary to ensure the positivity of $\lambda_\pm$ then read:
\begin{equation}
    c>0\,,\quad d\geq 0\,,\quad 1-\frac{4d}{c^2}\geq 0\,.
\end{equation}
The first two constraints above can be expressed as inequalities for quadratic polynomials in $\varepsilon^2$. 
For the benchmark model we find that upon setting
\begin{equation}\label{noghost}
    A-B<0\,,\quad A>0\,
\end{equation}
these are satisfied for any choice of $\varepsilon$, while the last condition is fulfilled automatically for arbitrary choice of $A,B,\varepsilon$.

The conditions necessary for avoiding the gradient instability are in turn
\begin{equation}\label{nograd}
    a>0\,,\quad b\geq 0\,,\quad 1-\frac{4bd}{a^2}\geq0\,
\end{equation}
and are slightly harder to satisfy. It is easy to see that by setting 
\begin{equation}
    A+B>1
\end{equation}
and assuming that \eqref{noghost} holds the condition $a>0$ can be satisfied for arbitrary values of $\varepsilon$. However, for these values of $A,B$ the equation $b=0$ defines an inverse parabola in the $\varepsilon^2$ space with two real roots $\varepsilon_{\pm}^2$ only when 
\begin{equation}
    B-1>0\,.
\end{equation}
Hence the condition $b\geq 0$ is only satisfied for $\varepsilon^2\in\left[\varepsilon_{-}^2,\varepsilon_{+}^2\right]$. Since we are only interested in positive values of $\varepsilon^2$ then we conclude that the condition $b\geq 0$ imposes a constraint on the maximal allowed strain applied to our system given by:
%\begin{equation}\label{emax}
%    \varepsilon_{\text{max}}^2=2\left[-1+\frac{\sqrt{(A%-B)(-1+A+B)-AB}}{\sqrt{A(A-B)}}\right]\,.
%\end{equation}
\begin{equation}\label{emax}
 %   \varepsilon_{\text{max}}^2=2\sqrt{2+\frac{B-1}{A}+\frac{A}{B-A}}-2\,.
     \varepsilon_{\text{max}}^2=2\sqrt{\frac{A(B-A)+A+B(B-1)}{A(B-A)}}-2\,.
\end{equation}
Analyzing the last condition in \eqref{nograd} analytically becomes more involved. We find however that in the parameter region 
\begin{equation}
    B\leq \frac{1}{2} \left(2-A+\sqrt{4-3 A^2}\right)\label{sep}
\end{equation}
the maximal strain is determined by the onset of the gradient instability and is thus given by \eqref{emax}. Only in the region complementary to \eqref{sep} is the maximal strain fixed by requiring the absence of superluminal propagation, finding
\begin{equation}\label{emaxSL}
    \varepsilon_{\text{max}}^2=2\sqrt{\frac{A (A+B-2)}{A^2+A (B-1)+(B-2) B}}-2\,.
\end{equation}
We present the full constraints on the parameter space obtained numerically in the main text.

Finally, let us quote our results for the simple case of linear deformations, \textit{i.e.} of zero background shear strain, \textit{i.e.} $\varepsilon =0$. We obtain the following allowed region for the exponents $A,B$:
\begin{equation}
    0\leq A\leq 1\quad \text{and} \quad  1\leq B\leq \sqrt{1-A}+1 \label{full2}\,.
\end{equation}
More specifically, the two kinetic eigenvalues in this case are equal and given by $\lambda_\pm=2^{-1+A-2B}B$ imposing the constraint $B>0$. The sound speeds are in turn given by $c_-^2=\frac{A}{B}$ and $c_+^2=B-1+\frac{A}{B}$. The absence of gradient instabilities is thus setting $A\geq0$ and $B-1\geq -A/B$. The latter constraint can be made stronger by requiring the positivity of the bulk modulus, leading to $B\geq1$; the positivity of the shear modulus gives again $A\geq 0$.\\

We can now repeat the exercise of finding the speeds and all the constraints for the nonrelativistic solid model presented in Section IV.
%The virtue of this potential can already be seen 
In the limit of infinitesimal strain, the transverse and longitudinal modes, as defined in \eqref{modes}, decouple at the level of the quadratic action \eqref{action2}. Indeed, for $\varepsilon = 0$ we find that $N_{TL}=c^2_{TL}=0$. From the positivity of the remaining quantities $N_T,N_L,c_T^2,c_L^2$ we arrive to the following set of conditions on the parameters of our model:
\begin{align}
A>0\,,\quad A+B(B-1)>0\,,\quad1+Bv^2>0\,.
\end{align}
The propagation speeds of the canonically normalized modes are then given as
\begin{align}
\frac{c_T^2}{N_T}=\frac{v^2A}{1+Bv^2}\,,\qquad\frac{c_L^2}{N_L}=\frac{v^2(A+B(B-1))}{1+Bv^2}\,.
\end{align}
We thus see that with this choice of potential both propagation speeds in the infinitesimal strain limit scale with $v$. Hence in order to go to non-relativistic speeds we just need to set $v\ll 1$. Let us also point out that the two speeds are related as $c_L^2=c_T^2+v^2B(B-1)$. The second term in this relation comes from the linear bulk modulus, defined in Eq.~\eqref{Klin}. For the new choice of potential it equals to $K=\rho_0v^2(B-1)B$ and thus for a negative $B-1$ the bulk modulus becomes negative. Henceforth we shall only consider $B\geq 1$.

The additional term in potential also enables us to expand the allowed parameter space for $A,B$. In particular, by analyzing the stability conditions \eqref{nograd} we find that the maximal strain is only set by the requirement of the absence of gradient instability for the parameter values $A<B$. Its value remains unaffected by the new term, \textit{i.e.} it does not depend on $v$, and is still given by \eqref{emax}, with the additional requirement (coming from $\varepsilon=0$) that $A+B(B-1)>0$. 
In the remaining the parameter space the maximal strain is determined by the superluminality constraint. The new term in the potential pushes the superluminality constraint further away thus expanding the allowed region for $A,B$.  This is shown in Fig.~\ref{emaxcase2}. \\

\bibliographystyle{apsrev4-1}

\bibliography{EFT}

\newpage

\mysection*{Erratum}

In Appendix \color{blue} A \color{black}, we computed the effective action for the phonon fluctuations  $\pi^I$ (see equations \color{blue}(A2) \color{black} and \color{blue}(A3)\color{black}). We have decomposed our fluctuations into transverse and longitudinal with respect to the wave-vector $\vec{k}$, rotated along the strain deformation of the equilibrium configuration. However, beyond linear level, for large strains, our background is not isotropic  because of the presence of an external strain configuration $\phi_{\rm str}^I$ which explicitly breaks the spatial SO(2) symmetry. Therefore, our original assumption that the field perturbations can only be made to depend on one of the spatial directions, i.e. that $\pi_{L/T}=\pi_{L/T}(t,x)$ appears unjustified and needs to be corrected. The nonlinear strain-dependent speed of propagation of the phonon fluctuations depends crucially on the angle of propagation $\theta$, which was considered to be only $\theta=0$ (longitudinal modes) and $\theta=\pi/2$ (transverse modes). In this erratum, we correct our previous results and we generalize them to a generic angle of propagation $\theta$ introduced as $\pi^I=a^Ie^{-i\omega t+ikx\sin\theta+iky\cos\theta}$. The main results remain unchanged. %\\[0.4cm]

The complete expression for the speeds $c_i^2(\theta,\varepsilon)$ is tedious but straightforward to compute from the quadratic action. %and can be found in the supplementary Mathematica notebook. 
The angle dependence disappears in the linear limit $c_i^2(\theta,\varepsilon\rightarrow 0)$ since the background becomes isotropic again. %\\[0.2cm]

%\begin{figure}
%\centering
%    \includegraphics[width=0.4\textwidth]{angle.pdf}
%    \caption{\MB{can we add a plot of $\varepsilon_{max}(\theta)$ to actually show which angle set the most stringent constraint and that such angle does not coincide with $0$ or $\pi/2$ as we used in our previous paper?}}
%    \label{fig:angle}
%\end{figure}

\begin{figure}[b]
    \centering
    \includegraphics[width=0.4\textwidth]{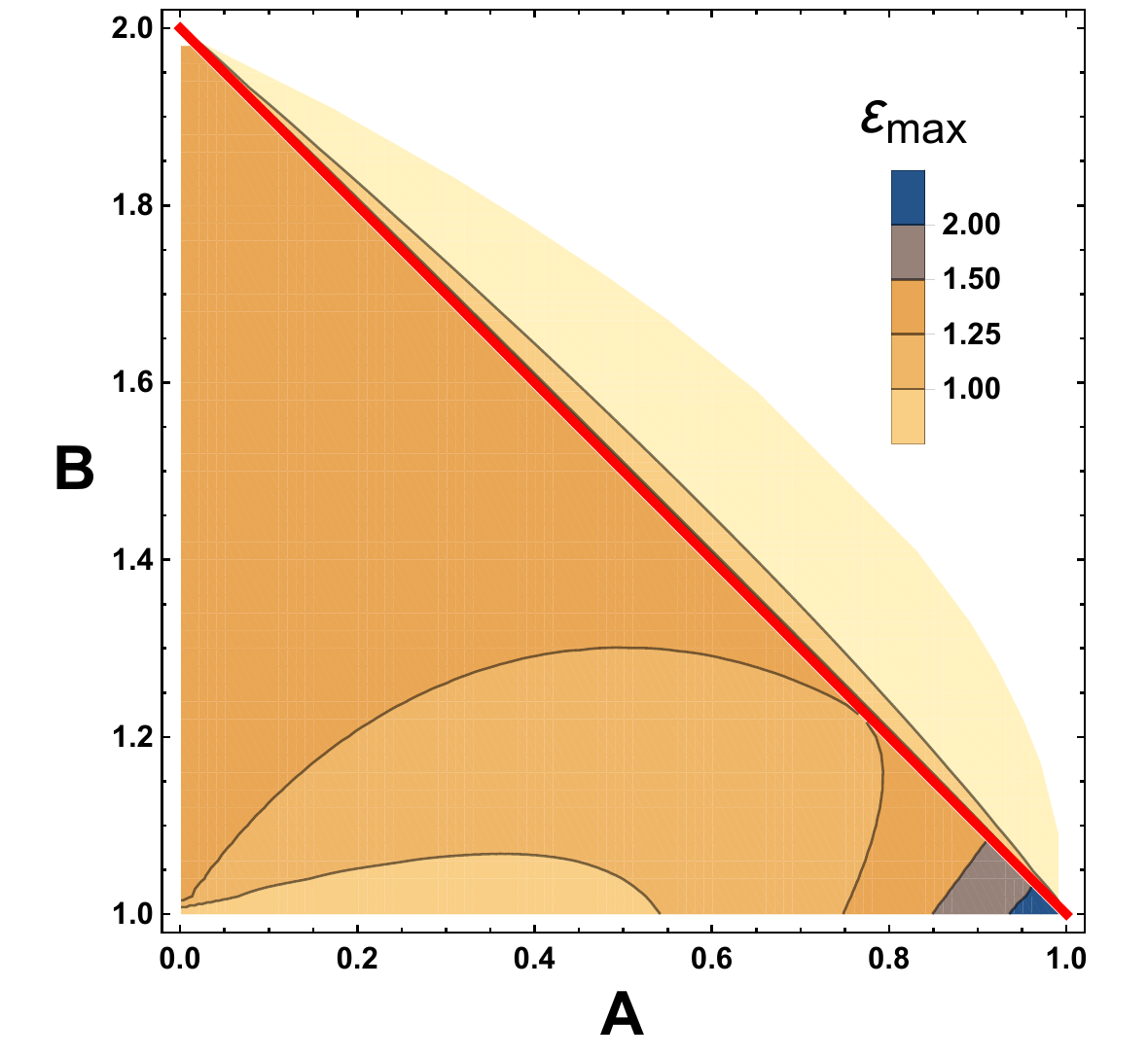}
    \caption{The allowed parameter region for the potential $V(X,Z) = \rho_{\rm eq}X^A Z^{\frac{B-A}{2}}$. The left, bottom and right edges are respectively given by: gradient instability, positivity of the bulk modulus, superluminality.  The red line separates the region where the maximum strain is due to the gradient instability (left) and the region where it is due to superluminality (right). Large strains are realized in the bluish area.}
    \label{fig:mono}
\end{figure}

Using the new expressions for the velocity and scanning all the angles $\theta$ we require that the same consistency requirements are satisfied in all directions: no ghosts, no gradient instabilities and no superluminal modes.\\
%\MB{In Fig.\ref{fig:angle} we show an example of how the maximal strain depends on the angle $\theta$ and how the most stringent constraint does not appear for $\theta=0,\pi/2$ as assumed in \cite{Alberte:2018doe}}. 
In Figures \ref{fig:mono}, \ref{emaxcase2} and \ref{emaxcase1} we find the maximum strain applicable $\varepsilon_{\rm max}$. In Fig.~\ref{fig:mono} we analyze the simplest of our benchmark models, $V(X,Z) = \rho_{\rm eq} X^A Z^{\frac{B-A}{2}}$, while in Figs.~\ref{emaxcase2} and \ref{emaxcase1} we work in the extended model with $V(X,Z)$ given in Eq.~(\textcolor{blue}{24}) (to be compared with Figures \color{blue}4\color{black}, \color{blue}5 \color{black} and \color{blue}6 \color{black}). The main difference that must be pointed out is that now we do not have anymore a region with large  $\varepsilon_{\rm max}$ close to $A=0$. Thus there are no hyperelastic materials with small shear exponent ($\nu_{\rm shear}=2A$)  anymore. Instead, the hyperelastic region at $A=B$ is not affected by the corrections. 
Notice that this last region $A=B$ is the one sharing important similarities with the nonlinear response of rubber materials \cite{enlighten70333}, as we explain in the Discussion section.  In this sense, the match with the phenomenology of hyperelastic materials is better with these new results.

\begin{figure}[t]
    \centering
    \includegraphics[width=0.4\textwidth]{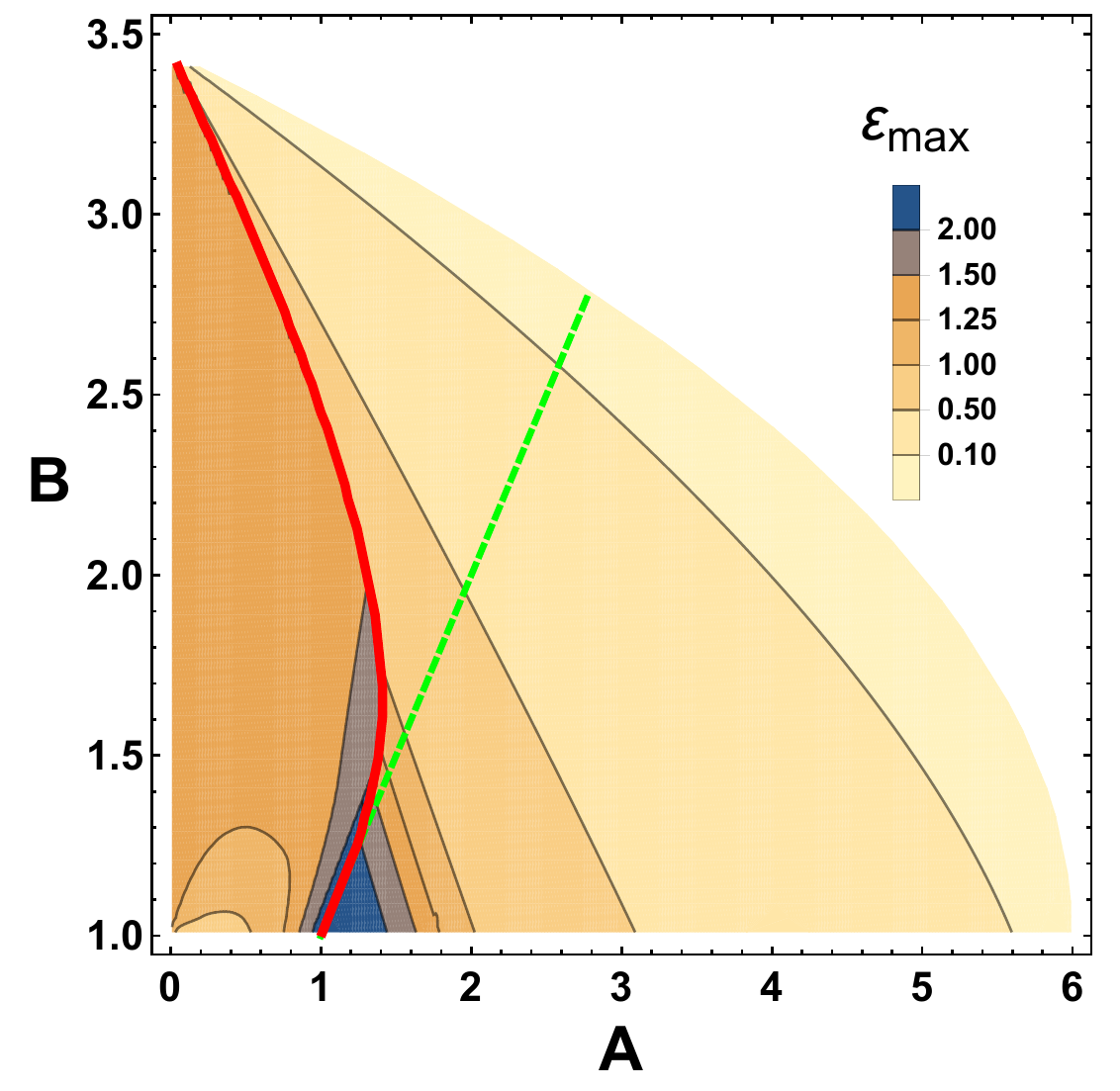}
    \caption{The expanded parameter space for $v^2 = 0.2$. The red line splits the regions where the limit on the maximal strain comes from superluminality (on the right) and from gradient instability (on the left). The green dashed line is $A=B$. In the region $A\geq B$ the maximum strain is only dictated by subluminality.}
    \label{emaxcase2}
\end{figure}

The full expression for $\varepsilon_{\rm max}$ as a function of $A,\,B$ 
is complicated but it simplifies in a certain region of the $A-B$ parameter space. Indeed, in the shaded area shown in Fig.~\ref{fig:analy1}, 
the form of $\varepsilon_{\rm max}$ is determined solely from avoidance of gradient instabilities ($c_i^2 > 0$) and reduces to the simple analytical form
\begin{equation}\label{analyticsol}
\varepsilon_{\rm max}=\sqrt{2}\,\left(\frac{(B-1)\,B}{A-A^2-B+B^2}\right)^{1/4}\,.
\end{equation}

%This result applies to both benchmark potentials as long as $c_i (\varepsilon_{\rm max}) < 1$, \LAi{in the opposite case $\varepsilon_{\rm max}$ will instead be determined by subluminality conditions. }

\begin{figure}[t]
    \centering
    \includegraphics[width=0.4\textwidth]{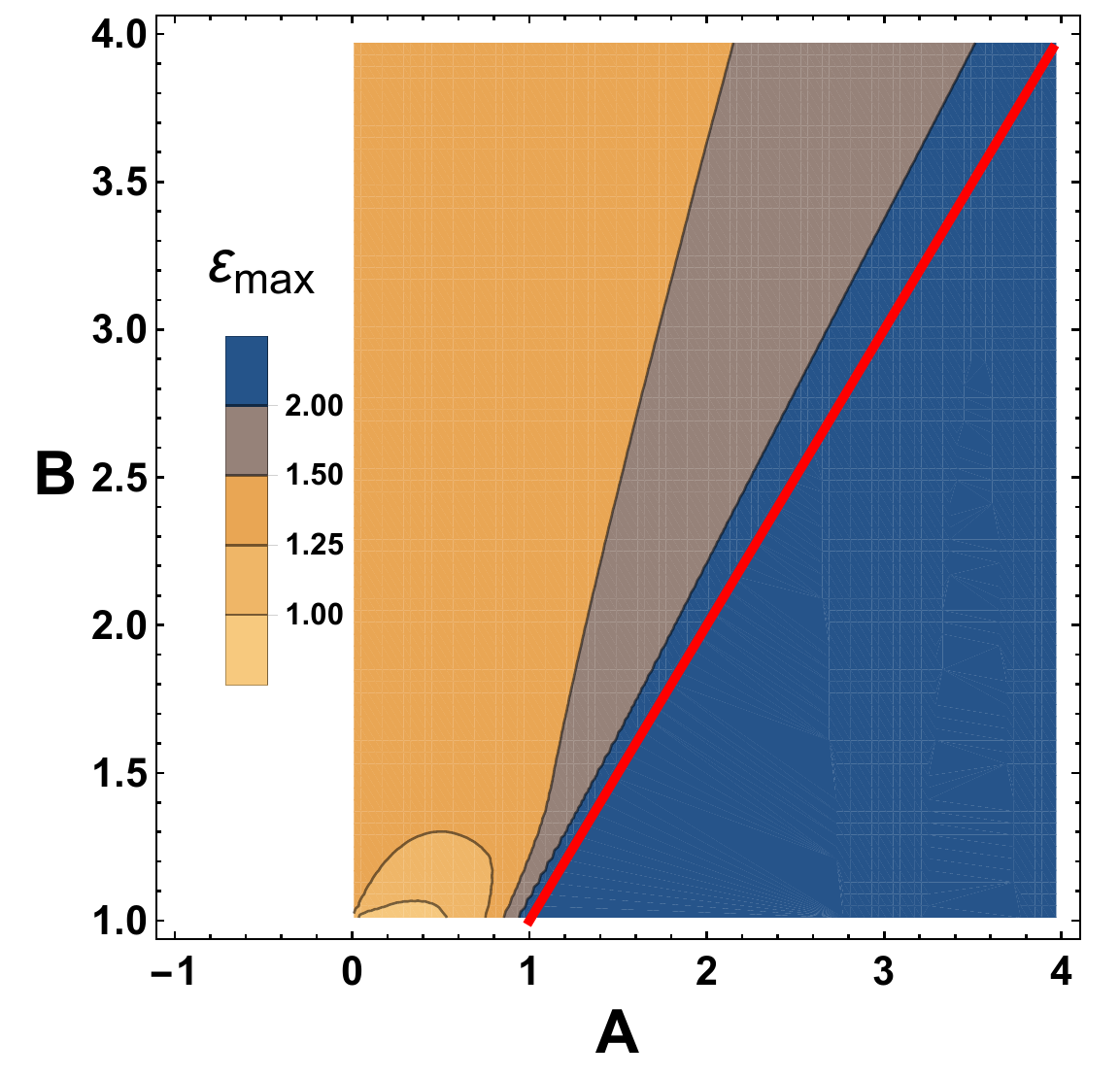}
    \caption{The expanded parameter space for $v^2 = 10^{-8}$. The subluminal constraint in $A<B$ is now located at larger values of $A$ and $B$. In the region $A\geq B$ the maximum strain is only dictated by subluminality.}
    \label{emaxcase1}
\end{figure}

\begin{figure}[h]
    \centering
    \includegraphics[width=0.46\textwidth]{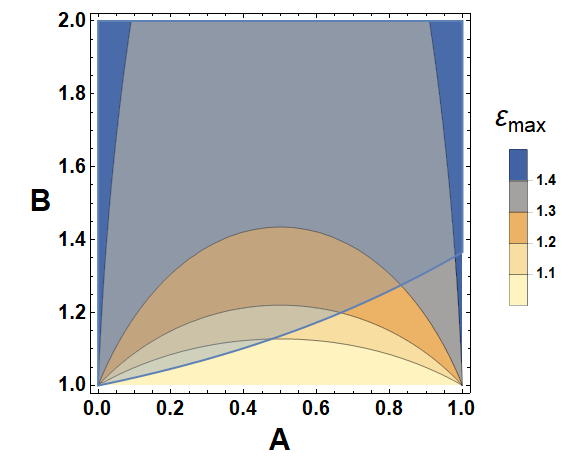}
    \caption{Analytic solution for $\varepsilon_{\rm max}$ without imposing subluminality constraints. The simple expression  \eqref{analyticsol} applies in the shaded area above the blue line.}
    \label{fig:analy1}
\end{figure}

In the main text we had found that, in the limit $A\rightarrow 0 $, $\varepsilon_{\rm max}$  increases asymptotically as
\begin{equation}\label{old}
\varepsilon_{\rm max} \simeq \sqrt{2}\,\left(\frac{B-1}{A}\right)^{1/4}\,.
\end{equation}
While this is correct for the cases $\theta = 0, \pi/2$ it is not true that this $\varepsilon_{\rm max}$ is valid for any generic value of $\theta$, so our original result was not the most restrictive one. In particular, from equation \eqref{analyticsol}, we observe that our previous interpretation of the above relationship \eqref{old} as a universal correlation between the maximum strain and exponents $A, B$ was not correct.  Instead, taking into account this angle dependence, we find an interesting universal limit
\begin{equation}
\varepsilon_{\rm max} \rightarrow \sqrt{2} \quad \mbox{for} \quad A \rightarrow 0\,.
\end{equation}
Let us remark that the value for $\varepsilon_{\rm max}$ close to $A=0$ is found for an angle $\theta$ that obeys $\sin (2\theta) = 1/\sqrt{3}$. For generic values of $A$ the value of $\theta$ leading to the most stringent condition on the maximal allowed strain has also been found analytically, however, its exact expression is too long to be given here. 

In the main text, we found another region where $\varepsilon_{\rm max}$ grows asymptotically, which is in the limit where $A\sim B$. This feature is still present, and it is given by the same expression up to a factor $2^{1/4}$:
\begin{equation}
\varepsilon_{\rm max} \sim \left(2\,\frac{A}{B-A}\right)^{1/4}
\end{equation}

\begin{figure}[b]
    \centering
    \includegraphics[width=0.4\textwidth]{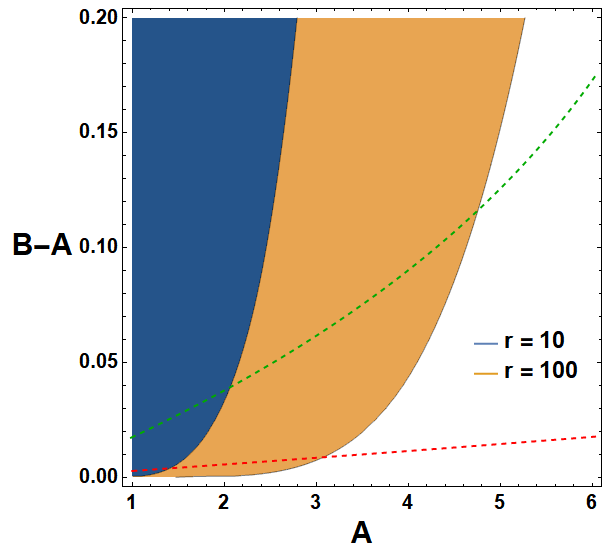}
    \caption{Speed ratio constraints for $r=10,10^2$. Dashed lines show where $\varepsilon_{\rm max}=3,5$ (Green, Red).}
    \label{fig:ratio}
\end{figure}

Finally we would like to revisit the ratio considered for the non-relativistic potential between the speed $c_+$ at zero strain and at the maximum shear deformation
\begin{equation}
r\equiv \frac{c_+(\varepsilon_{\rm max})}{c_+(\varepsilon =0)}\,,
\end{equation}
which is shown in Figure \ref{fig:ratio}. Comparing with the one obtained originally we see that the parameter space has decreased only slightly. Thus, we continue to have a finite region where the $\varepsilon_{\rm max}$ is large  and $r$ is moderately large (the material is non-relativistic throughout the deformation). Interestingly, this includes values of the exponents $\nu_{\rm shear}, \,\nu_{\rm bulk}$ of order of a few --  as seen in some real-world rubber-like materials \cite{enlighten70333}. 

\end{document}